\documentstyle[preprint,prb,aps]{revtex}
\tighten
 
\begin{document}
\draft
\title{Fluctuation--induced phase in CsCuCl$_3$ in transverse magnetic field: 
Theory}
 
\author{A. E. Jacobs\cite{Allan} and Tetsuro Nikuni\cite{Tetsuro}}

\address{Department of Physics, University of Toronto,
Toronto, Ontario, Canada M5S~1A7}
 
\date{\today}

\maketitle
 
\begin{abstract}

CsCuCl$_3$ is a quantum triangular antiferromagnet, ferromagnetically 
stacked, with an incommensurate (IC) structure due to a 
Dzyaloshinskii--Moriya interaction. 
Because of the classical degeneracy caused by the frustration, fluctuations 
in CsCuCl$_3$ have extraordinarily large effects, such as the phase transition 
in longitudinal magnetic field (normal to the planes, parallel to the IC 
wavenumber ${\bf q}$) and the plateau in ${\bf q}$ in transverse field 
(perpendicular to ${\bf q}$). 
We argue that fluctuations are responsible also for the new IC phase 
discovered in transverse field near the N\'{e}el temperature $T_{\rm N}$, by 
T. Werner {\it et al}. [Solid State Commun. {\bf102}, 609 (1997)]. 
We develop and analyse the corresponding minimal Landau theory; 
the effects of fluctuations on the frustration are included 
phenomenologically, by means of a biquadratic term. 
The Landau theory gives two IC phases, one familiar from previous studies; 
properties of the new IC phase, which occupies a pocket of the 
temperature--field phase diagram near $T_{\rm N}$, agree qualitatively with 
those of the new phase found experimentally. 

\end{abstract}
 
 
\section{Introduction}

Compounds based on the triangular antiferromagnet (TAFM), particularly 
members of the $ABX_3$ family, have provided a wealth of interesting 
behaviour and indeed many surprises, as recently reviewed\cite{collins97}. 
Because of the extraordinarily large effects of quantum and thermal 
fluctuations, CsCuCl$_3$ (with N\'{e}el temperature\cite{achiwa69} 
$T_{\rm N}=10.65\,$K and zero--temperature saturation field 
$H_{\rm S}\approx30\,$T) ranks among the most interesting of these compounds; 
the surprises began 20 years ago\cite{moto78} and continue still,
in theory as well as in experiment. 

The magnetic properties of CsCuCl$_3$ are due to the Cu$^{2+}$ ions 
($S=1/2$); these form a stacked triangular lattice, to a good approximation. 
The interaction in the planes is antiferromagnetic and therefore frustrated. 
Normal to the planes (in the chain or $c$ direction), a 
ferromagnetic interaction competes with a Dzyaloshinskii--Moriya (DM) 
interaction\cite{DM}, giving an incommensurate (IC) structure 
with wavenumber ${\bf q}=q\hat{\bf z}$ in the $c$ direction. 
In more detail, the classical, zero--temperature structure in zero magnetic 
field is the three--sublattice, $\pm120^\circ$ TAFM structure; the spins 
lie in the planes, rotating by $\approx5.1^\circ$ per plane\cite{adachi80}.  
Application of a magnetic field yields a variety of interesting phenomena 
related to the classical degeneracy of the TAFM; 
recall that the classical ground state of the TAFM is continuously (and also 
discretely) degenerate, even in a magnetic field, and also that thermal 
fluctuations\cite{lee84,kawa84,kawamiya84,korsh86} in classical models and 
quantum fluctuations\cite{nishi86,chub91} break the continuous degeneracy in 
the same way, both selecting for example the colinear structure at 
$H\approx H_{\rm S}/3$. 

In longitudinal magnetic field (normal to the planes, parallel to the IC 
wavenumber ${\bf q}$), the low--temperature magnetization is 
discontinuous\cite{moto78,fed85,nojiri88} at $H\approx0.4H_{\rm S}$, 
due to a novel, fluctuation--induced phase transition\cite{shiba93,nikuni93}: 
the umbrella structure is optimal at small $H$ (due to a small, easy--plane 
anisotropy in the interplane exchange\cite{tanaka92}) and a coplanar 
structure is optimal at larger $H$ (due to quantum fluctuations). Other 
experiments\cite{ohta93,chiba94,mino94,ohta94,schotte94,chiba95,moto95,weber96} 
support the Nikuni--Shiba analysis\cite{shiba93,nikuni93}. 
In summary, CsCuCl$_3$ in longitudinal field appears to be understood, except 
that the transition at high temperature $T$ has puzzling 
features\cite{weber96}. 

A transverse field (in the planes, perpendicular to ${\bf q}$) gives more 
surprises. 
The behaviour at low fields is conventional: $q$ decreases quadratically with 
$H$, and the curvature increases\cite{mino94,schotte94} with $T$. 
Classical (mean--field) theory\cite{jacobs93,ohyama95} gives both of these 
and also a transition to the commensurate (C) phase at $H\approx0.47H_{\rm S}$ 
(at $T=0$); 
this transition has recently been observed\cite{nojiri97}, but at larger 
$H$ ($\approx0.58H_{\rm S}$).  
Classical theory cannot, however, explain any of the following: 
At low $T$, unusual behaviour occurs for fields near $H=H_{\rm S}/3$, 
well below the IC$\to$C transition; structure is found 
in the magnetization\cite{nojiri88}, 
in the $^{133}$Cs NMR shift\cite{chiba95}, 
in the IC wavenumber (Ref.\onlinecite{mino94} finds a plateau), 
and in ESR measurements\cite{ohta94}. 
At high $T$, a second IC phase appears and $T_{\rm N}$ increases with 
field\cite{werner97}. 

At low $T$, the structure near $H_{\rm S}/3$ seems related to changes, 
induced by quantum fluctuations, in the structure of the TAFM near the same 
field\cite{nishi86,chub91}. 
Linear spin--wave (LSW) theory, which adds the leading quantum correction 
to classical theory, gives a plateau in the magnetization of the C 
state\cite{nikuni98} (as for the TAFM), a promising start. 
But there is another surprise, this time in theory. 
Not only does LSW theory of the IC phase fail to explain any of the other 
results, it actually provides a worse description of the IC phase 
than does classical theory, by predicting a premature IC$\to$C 
transition\cite{nikuni98} at $H\approx0.32H_{\rm S}$. 
An innovative phenomenological theory\cite{nikuni98} of quantum fluctuations 
explains the existence of the plateau\cite{mino94,nojiri97} in $q$, and also 
its level. 
The theoretical value ($H\approx0.44H_{\rm S}$) for the field at the 
IC$\to$C transition\cite{nojiri97} is however too small, and the 
magnetization is not predicted well; likely the phenomenological theory 
can be improved. 

Both new findings\cite{werner97} near $T_{\rm N}$, namely the increase of 
$T_{\rm N}$ with field (as in the TAFM\cite{lee84,kawamiya84,miya86JPSJ}) and 
the second IC phase, are likely due to thermal fluctuations; 
neither has yet been treated theoretically. 
Because a microscopic or numerical treatment of fluctuations is out of the 
question for a vector--spin system with a nonsinusoidal IC structure, we 
use phenomenology. 
To treat a particular aspect of fluctuations near $T_{\rm N}$, namely effects 
related to the breaking of the classical degeneracy of the TAFM, we add to 
the standard Landau theory a term biquadratic in the order parameter, as in 
our treatment\cite{nikuni98} of quantum fluctuations at $T=0$; this term 
appears neither in the Hamiltonian nor in mean--field theory\cite{ohyama95} 
at any $T$. 
Of course this term is not intended to include fluctuation effects in general 
(such phenomenology cannot possibly explain the reduction of $T_{\rm N}$ from 
the mean--field value of $35.5\,$K to the experimental value of $10.65\,$K). 

In qualitative agreement with experiment\cite{werner97}, the Landau theory 
predicts a second IC phase to exist in a small $T$--$H$ region near 
$T_{\rm N}$. 
As discussed in the preceding article\cite{schotte98} and in Section IV, the 
theory explains other properties of the new phase, qualitatively; 
as discussed in the Appendix, however, it does not explain the increase of 
$T_{\rm N}$ with $H$. 
The new phase is the high--$T$ version of a state which arose in the 
classical theory\cite{jacobs93}; the state exists at all $T$ in classical 
(mean--field) theory\cite{jacobs93,ohyama95}, but is never optimal. 
We argue that the new phase owes its existence to fluctuations; 
these are strong enough to overcome an energy difference in classical theory, 
just as quantum fluctuations in longitudinal field overcome the small 
anisotropy\cite{shiba93,nikuni93}. 
A unifying feature is that the two most striking of the experimental results 
in transverse field, namely the plateau\cite{mino94,nojiri97} in $q$ and the 
new phase\cite{werner97} near $T_{\rm N}$, are explained using the same 
phenomenological treatment of fluctuations, in Ref.\onlinecite{nikuni98} 
and here respectively. 

Remarkably then, CsCuCl$_3$ displays a fluctuation--induced phase transition 
in transverse field, and a different fluctuation--induced phase transition 
in longitudinal field\cite{shiba93,nikuni93}. 

\section{Hamiltonian}

The main interactions are described by the Hamiltonian 
$${\cal H}=\sum_{in}\Bigl[\,-\,2\,J_0\,{\bf S}_{in}\cdot{\bf S}_{i,n+1}
-D\,\hat{\bf z}\cdot({\bf S}_{in}\times{\bf S}_{i,n+1})
+\,J_1\mathop{{\sum}'}_k{\bf S}_{in}\cdot{\bf S}_{kn}
-g\,\mu_{\rm B}\,H\,\hat{{\bf x}}\cdot{\bf S}_{in}\,\Bigr]\ ,\eqno(1)$$
where ${\bf S}_{in}$ is the spin operator at the $i$th site in the 
$n$th $a$--$b$ plane, $\hat{\bf z}$ and $\hat{\bf x}$ are unit vectors in 
the $c$ and $a$ directions, and the $k$ sum is over the six, in--plane, 
nearest neighbours of the site $in$. 
The first term ($\propto J_0$) is the isotropic, ferromagnetic exchange 
interaction between spins in nearest--neighbour planes, 
the second ($\propto D$) is the interplane DM interaction, 
the third ($\propto J_1$) is the isotropic, frustrated, antiferromagnetic 
exchange interaction between nearest-neighbour spins in the $a$--$b$ planes, 
and the fourth is the Zeeman energy in a field ${\bf H}$ transverse to the 
chains. 
We omit the easy--plane anisotropy\cite{tanaka92} in the interplane 
interaction, the dipole--dipole interaction, and several other effects. 
The coefficients have been estimated 
previously\cite{achiwa69,adachi80,hyodo81,tazuke81,tanaka92,mekata95}; 
we use $J_0=28\,$K, $J_1=4.9\,$K and $D=5\,$K. 
The saturation field, above which each spin is aligned with the 
field at $T=0$, is $H_{\rm S}=18J_1S/(g\mu_{\rm B})\approx 30\,$T. 

At the classical level, the intrachain exchange term $(J_0$) favours states 
with spins parallel in adjacent $a$--$b$ planes while the smaller DM term 
($D$) favours states with spins in the planes and rotating by $\pi/2$ per 
plane. 
At zero field, for all $T<T_{\rm N}$, the spins lie in the planes, forming 
the $120^\circ$ structure with three sublattices. 
The structure normal to the planes is helical; the wavenumber at $H=0$ is 
$q_0\,\hat{\bf z}$ where $q_0=\arctan\left(D/(2J_0)\right) \approx2\pi/71$. 
A transverse field deforms the helical structure, which becomes highly 
nonsinusoidal at higher $H$. 

For $T=0$, the above Hamiltonian was investigated in the classical 
approximation\cite{jacobs93}, and the leading quantum correction was 
obtained using linear spin--wave (LSW) theory\cite{nikuni98}. 
Neither theory can account for the structure observed near $H_{\rm S}/3$, 
but a phenomenological treatment\cite{nikuni98} of quantum fluctuations 
is largely successful. 

The extension of classical theory to $T>0$ (by mean--field theory) gives a 
phase diagram\cite{ohyama95} with only one IC phase, and so an understanding 
of the new IC phase near $T_{\rm N}$ seems to require including fluctuations 
at some level. 
A satisfactory microscopic treatment of fluctuations in CsCuCl$_3$ is out of 
the question at $T=0$, and they are even more difficult to treat for $T>0$, 
leaving it seems only a phenomenological approach. 
For general $T>0$, one could simply add the biquadratic term\cite{nikuni98} 
to the mean--field expression\cite{ohyama95} for the free energy, but the 
coefficient would have to be adjusted as fluctuation effects increase with 
$T$, requiring a fit at each $T$ or strong guidance from theory; 
this approach would be most reasonable near $T_{\rm N}$, if the mean--field 
free energy were expanded to fourth order in the order parameter. 
We have chosen instead to use a fourth--order Landau theory. 

\section{Landau theory}

The following describes a minimal Landau theory of CsCuCl$_3$ near 
$T_{\rm N}$; 
usually, Landau theory is reliable regarding the phase diagram, less reliable 
regarding the order of the transitions, and unreliable regarding fine 
details like the position dependence of the order parameter. 
We assume a structure with three sublattices in the $a$--$b$ planes and 
with period $L$ in the $c$ direction (in units of the layer spacing); 
the restriction to integer $L$ causes no difficulty\cite{nikuni98}. 
We assume also that the spins remain in the $a$--$b$ planes at all $H$ and 
$T$. 
Curiously, the DM term is not sufficient for this\cite{jensen,nikuni98}; 
the easy--plane anisotropy\cite{tanaka92} helps of course, but fluctuation 
effects seem to be necessary\cite{nikuni98}. 
In mean--field theory, the free energy would be expressed in terms of the 
site--dependent magnetization $\langle{\bf S}_{jl}\rangle$, where 
$j=1,2,3$ is the sublattice index and $l=1,\ldots,L$ is the layer index. 
In Landau theory, the free energy is expanded in the order parameter 
${\bf m}_{jl}$, which is only proportional to $\langle{\bf S}_{jl}\rangle$. 

Explicitly, we use the following expression for the free energy $F$ of the 
$N$ spins, relative to the paramagnetic state at $H=0$: 
$$F={N\over{3L}}\sum_{j=1}^3\sum_{l=1}^L \Bigl[\,
{\textstyle{1\over2}}\,\alpha_1\,{\bf m}_{jl}^2
+         \alpha_2\,{\bf m}_{jl}\cdot{\bf m}_{j+1,l}
      - h\,\hat{{\bf x}}\cdot{\bf m}_{jl}
+{\textstyle{1\over4}}\,\gamma_1\,{\bf m}_{jl}^4
-{\textstyle{1\over2}}\,\gamma_2
\left({\bf m}_{jl}\cdot{\bf m}_{j+1,l}\right)^2 $$
$$ -\delta_1\,{\bf m}_{jl}\cdot\left({\bf m}_{j,l+1}-{\bf m}_{jl}\right)
-\delta_2\,{\hat{\bf z}}\cdot\left( {\bf m}_{jl}\times {\bf m}_{j,l+1}\right)
\,\Bigr] \ . \eqno(2)$$
This expression differs in major respects from the mean--field expansion 
to fourth order in ${\bf m}$ (or $\langle{\bf S}\rangle$). 
On the one hand, we omit from the expansion many terms of the same order as 
the ones we keep, second--order terms as well as fourth--order terms;  
on the other, and more importantly, we add a fourth--order term which does not 
appear in the expansion. 
All seven terms are essential. 
The third term in Equation (2) is the Zeeman energy; 
the Landau parameter $h$ is proportional to the magnetic field $H$. 
The terms with coefficients $\alpha_1$ and $\gamma_1$ are standard, while 
those with coefficients $\alpha_2$, $\delta_1$ and $\delta_2$ result 
respectively from the in--plane antiferromagnetic interaction, the 
interplane ferromagnetic interaction and the DM interaction. 
The terms in $\alpha_1$ and $\delta_1$ are adjusted so that the latter makes 
no contribution to the energy of the commensurate state. 

The remaining term, the biquadratic term 
$-{1\over2}\gamma_2\left({\bf m}_{jl}\cdot{\bf m}_{j+1,l}\right)^2$, appears 
neither in the Hamiltonian nor in the mean--field theory\cite{ohyama95}. 
It is introduced purely phenomenologically, to represent (to the extent 
possible) the effect of fluctuations (here thermal); 
the first such phenomenological use of the term was in 
Ref.\onlinecite{nikuni98}. 
In the TAFM, fluctuations (quantum or thermal) act to favour the colinear 
phase at intermediate fields, and therefore we require $\gamma_2>0$; 
since the strength of the fluctuations is however unknown, $\gamma_2$ is 
otherwise adjustable (except that $\gamma_2<0.5$ for stability). 
In the TAFM, and in the C state of CsCuCl$_3$, the fluctuations break the 
classical degeneracy, selecting one state from the many. 
Their effect in the IC phase of CsCuCl$_3$ is far more subtle, for two 
reasons: first, the IC structure forbids the quantum selection possible 
in a C structure; second, the classical IC phase is unconventional, being 
well described as a continuous sequence of degenerate commensurate 
states\cite{jacobs93}. 
The fluctuations act to reorient the spins, giving a conventional IC 
phase with domain walls separating nearly commensurate regions. 
This is why LSW theory of the IC phase fails: a treatment beyond LSW 
theory is necessary to account for the reorientation. 

The above Landau theory is of course related to the mean--field theory of the 
ferromagnetically stacked TAFM near $T_{\rm N}$. 
Equation (3) of Ref.\onlinecite{plumer94} (on the $XY$ model), with 
$B_i=9T_{\rm N}/5$ for $i=1$ to 6, reduces to our Equation (2) with the last 
three terms (coefficients $\gamma_2$, $\delta_1$ and $\delta_2$) omitted, 
although some effort is needed to see this. 
As discussed in Ref.\onlinecite{plumer94}, the colinear phase stable at 
intermediate fields cannot be obtained from this theory; 
it is obtained however if only $B_4$ is different from (less than) the other 
$B_i$, or if $B_2$ (or $B_3$) is different. 
Ref.\onlinecite{plumer94} noted that some effects of fluctuations can be 
mimicked by allowing some Landau coefficients to depart from their mean--field 
values; these authors did not consider the biquadratic term explicitly, but 
their intention was like ours, to break the classical TAFM degeneracy. 

From the Hamiltonian, we expect the coefficients $\alpha_2$, $\delta_1$ and 
$\delta_2$ to be proportional to $6J_1$, $2J_0$ and $D$ respectively. 
To reduce the number of parameters, we normalize the order parameter and the 
free energy so that $\alpha_2=1$ and $\gamma_1=1$. 
Then the interlayer coefficients are $\delta_1=2J_0/(6J_1)\approx1.9$ and 
$\delta_2=D/(6J_1)\approx0.17$. 
Of the remaining parameters $\alpha_1$, $h$ and $\gamma_2$, 
only $\alpha_1$ depends on $T$. 
At $H=0$, the IC--paramagnetic transition at $T_{\rm N}(0)$ occurs at 
$\alpha_1\approx1.015$; 
$\alpha_1=1$ is the upper limit of the C phase at $H=0$. 
In Landau theory, the transition at $T_{\rm N}(0)$ is second--order. 
Experiment\cite{weber96} finds tricritical or weakly first--order behaviour; 
the latter is obtained in a recent Monte--Carlo analysis\cite{plumer97} of 
a related model. 

The unknown constants of proportionality involving ${\bf m}$ and $h$ can be 
determined by comparing the Landau and mean-field theories. 
From Equation (2), the Landau energy (per spin) of the paramagnetic state 
at $\alpha_1=1$ is ${3\over2}m^2 -hm+{1\over4}m^4\ .$ 
The corresponding mean--field expression is found by setting 
$\beta\,S^2\,J_{\rm C}=1$ in Equation (3.2) of Ref.\onlinecite{ohyama95}: 
$${{\cal J}\over{J_{\rm C}}}\left( 
18\,J_1\langle S\rangle^2
-2\,g\,\mu_{\rm B}\,H\langle S\rangle
+{{4}\over{3S^2}}{{{\cal J}^3}\over{J_{\rm C}^2}}\langle S\rangle^4
+\cdots\right)\ ,$$ 
where $J_{\rm C}=4\,J_0+6\,J_1$ and ${\cal J}=2\,J_0-6\,J_1$. 
Because the mean--field expression omits fluctuations, the term in 
$\gamma_2$ must be omitted from the Landau expression. 
On setting $\langle S\rangle=am$ and comparing coefficients, one finds 
$h=k\,H/H_{\rm S}(0)$ where $k=3S/a$ and $k^2=4{\cal J}^3/(J_{\rm C}^2J_1)$; 
the numerical value is $k\approx0.88$. 

The Euler--Lagrange equations of the above Landau theory have many solutions, 
namely the paramagnetic (P) solution ${\bf m}_{jl}=m_{\rm P}\hat{\bf x}$, 
several commensurate (C) solutions, and many incommensurate (IC) solutions; 
the following provides some background for the last. 
In classical theory\cite{jacobs93}, the magnitudes of the site magnetizations 
are fixed and the phases suffice for a complete description. 
A single IC solution is optimal at all fields below the IC$\to$C transition 
(predicted to occur at $H\approx0.47H_{\rm S}$); 
in this solution (called the 111 solution), the spins on all three sublattices 
wind through $2\pi$ over a period $L$ of the IC structure. 
Many other IC solutions exist at fields below the transition. 
In the 110 solution, which is never optimal, the spins on only sublattices 
1 and 2 wind through $2\pi$ over one period, while the spins on sublattice 3 
wobble about the field without winding. 
The 111 and 110 solutions, and many other solutions generated from them by 
forming composite solutions\cite{jacobs93}, become degenerate at the 
IC$\to$C transition, which is therefore a multi--phase point. 
In mean--field theory\cite{ohyama95}, for $T>0$, the magnitudes of the 
magnetizations are no longer fixed; they can adjust to minimize the energy 
(for example by decreasing near a domain wall). 
The solution corresponding to the 111 solution is again optimal in all cases; 
it loses its winding character at larger $T$ and $H$ where the orbit in the
$m_x$--$m_y$ plane no longer encircles the origin. 
The infinite degeneracy at the IC$\to$C transition remains at $T>0$. 

The Landau--theory states corresponding to the 111 and 110 solutions are most 
easily described at $H=0$ (where the wavenumber is $q_0$ for both). 
For the first (IC$_1$), 
$${\bf m}_{jl}=m_1\left[\hat{\bf x}\,\cos(q_0l+\phi_j)
                       +\hat{\bf y}\,\sin(q_0l+\phi_j)\right]
\ \ \ \ {\rm for} \ j=1,2,3\eqno(3)$$ 
with $\phi_j=\phi_0+(j-1)2\pi/3$. 
For the second (IC$_2$), 
$${\bf m}_{1l}=-{\bf m}_{2l}=m_2\left[\hat{\bf x}\,\cos(q_0l+\phi_0)
                                     +\hat{\bf y}\,\sin(q_0l+\phi_0)\right]
\ \ \ ,\ \ \ {\bf m}_{3l}=0\ ;\eqno(4)$$
this is stable at $H=0$ only for $\alpha_1\agt0.99$, but solutions 
at lower $\alpha_1$ and $H>0$ are easily found. 
The first has the lower energy at $H=0$, for all $T$. 
For the same amplitude, the first optimizes the $\alpha_2$ term in the 
density of Equation (2), while the second optimizes the biquadratic term; 
the respective energies are $-{3\over2}\alpha_2m^2-{3\over8}\gamma_2m^4$ and 
$-\alpha_2m^2-{1\over2}\gamma_2m^4$. 
It is then possible that the second can have the lower energy, though only 
for $H>0$. 

For $H>0$, some analytical results can be found at small 
field\cite{jacobs93,ohyama95}, but full numerical solutions of the 
Euler--Lagrange equations are required at general values of $H$. 
Solutions were found by repeated linearization about trial solutions 
and solution of the linearized equations for the corrections. 
With increasing $H$, both solutions evolve, becoming increasingly 
nonsinusoidal; other solutions are found, but these are never optimal. 
In the second solution, the order parameter on the third sublattice 
increases from 0 and wobbles about ${\bf H}$ with period $L/2$. 
Phase diagrams follow from comparison of the energies of the IC, P and 
C solutions. 

\section{phase diagram and other results}

Because Equation (2) cannot explain both the new IC phase and the increase of 
$T_{\rm N}$ with $H$, and also because Equation (2) omits many terms of the 
same order as those kept, a detailed comparison with experiment is not 
attempted; fine details of the results should not be taken seriously. 
The only parameter in the theory (apart from the unknown constant 
relating $\alpha_1$ to $T$), is the coefficient $\gamma_2$ which determines 
the strength of the fluctuations. 
The value $\gamma_2=0.2$ used in the following was chosen, with guidance from 
experiment\cite{schotte98}, to give a reasonable size to the IC$_2$ region of 
the phase diagram, and also to the plateau in $q/q_0$ for the IC$_1$ phase 
(as a function of $h$); $0.1$ seems too small and $0.3$ too large. 
Actually, the new phase appears even at $\gamma_2=0$, but only in a thin 
sliver ($\Delta h\leq0.024$) of the phase diagram, with re--entrance and 
with no sizeable plateau. 

Figure 1 shows the theoretical phase diagram near $T_{\rm N}$; 
again, $\alpha_1$ is linear in $T$ [$\alpha_1=1.015$ at $T_{\rm N}(H=0)$] 
and $h\approx0.88 H/H_{\rm S}(0)$. 
The two IC phases (IC$_1$ and IC$_2$), the P phase, and the C phase are 
optimal in the regions indicated; 
the C phase is the ${\bf m}_1={\bf m}_2$ phase of the stacked TAFM. 
The new feature here is the IC$_2$ phase. 

The companion article\cite{schotte98} presents the strongest evidence for 
identifying the IC$_2$ phase with the new IC phase discovered in 
Ref.\onlinecite{werner97}: 
the neutron--scattering intensity as a function of wavenumber is 
qualitatively that expected from the order parameter of the IC$_2$ phase.  
The companion article compares theory and experiment in other respects as well. 
The following compares several aspects of the phase diagrams; 
it also presents results for the dependence of the IC wavenumbers on $T$ 
and $H$, and for the order parameters in the two IC phases. 

Phase diagrams (qualitative aspects): 
In both Figure 1 and Figure 3 of Ref.\onlinecite{schotte98}, the new phase 
appears in a small $T$--$H$ region near $T_{\rm N}$. 
Landau theory misses the increase of $T_{\rm N}$ with $H$ and therefore also 
the nose; experimentally, the region of the new phase is shaped more like a 
croissant than a pocket. 
In both theory and experiment, the new phase appears only above some field, 
and a narrow tail extends to the low--$T$ side. 

Phase diagrams (order of the transitions): 
The fields available in the high--$T$ neutron 
measurements\cite{werner97,schotte98} were not sufficient to observe the 
C phase, nor of course the IC$_1$$\to$C and C$\to$P transitions; 
the IC$_1$$\to$C transition was however observed\cite{nojiri97} at low $T$, 
at $H\approx18\,$T. 

\noindent 
In theory, all four of the other transitions in Figure 1 are first--order. 
IC$_1$$\to$IC$_2$ is strongly first--order, while the three transitions 
to the P state 
(IC$_1$$\to$P near $T_{\rm N}$, IC$_2$$\to$P and IC$_1$$\to$P at lower $T$) 
are weakly first--order (the free energies cross with almost the same slope); 
the IC$_1$$\to$P transition is second--order at $H=0$. 

\noindent 
In experiment, only IC$_1$$\to$IC$_2$ is unambiguously first--order. 
The scan\cite{schotte98} at $10.34\,$K can be interpreted in two 
ways\cite{schotte98b}: either the tail of the IC$_2$ phase was missed, or 
there is a first--order IC$_1$$\to$P transition at $H\approx12\,$T. 
The IC$_2$$\to$P transition is almost certainly second--order, from the 
observation of critical scattering\cite{schotte98}. 

Phase diagrams (other aspects): 
In theory, the IC$_2$ phase is not found at any $T$ if $h<0.10$ or if $h>0.34$, 
corresponding to $H<3.4\,$T and $H>11.6\,$T. 
In experiment, the lower limit is $H=4.3\pm0.3\,$T; if the IC$_2$ phase does 
not appear at $10.34\,$K, then the upper limit is $11.5\pm0.5\,$T. 
The agreement is reasonable. 
Independent of the constant $k$ relating $h$ and $H$, the relative widths 
in the field variable are comparable: $11.5/4.3\approx0.34/0.10$. 
The nose in the experimental phase diagram prevents a similar analysis 
for the temperature variable; 
for example, we cannot estimate reliably the upper $T$ limit of the 
C phase (but none of our estimates disagrees with the data). 

Wavenumbers of the IC phases: 
Figure 2 shows theoretical results for the reduced wavenumber $q/q_0$ as a 
function of $h$, at four values of $\alpha_1$. 

\noindent For $\alpha_1=1.005$, there are two transitions as $h$ increases, 
IC$_1$$\to$IC$_2$ at $h\approx0.137$ and IC$_2$$\to$P at $h\approx0.202$, 
both first--order in theory; 
at the first, $q$ increases discontinuously to a value less than the 
zero--field value $q_0$. 
In theory, $q/q_0$ for the IC$_2$ phase is roughly independent of $h$, 
for fixed $\alpha_1$. 
The dependence on $\alpha_1$ is stronger, but still weak; 
the value decreases roughly linearly with $\alpha_1$, 
from $q/q_0\approx0.98$ at $\alpha_1=1.01$ 
  to $q/q_0\approx0.88$ at $\alpha_1=0.995$. 
The experimental value\cite{schotte98} for $q/q_0$ in the IC$_2$ phase is 
$\approx0.87$; this is larger than in the IC$_1$ phase, as in theory; 
on the other hand, no dependence on $T$ or $H$ was observed. 

\noindent 
At lower $T$ ($\alpha_1=0.98$ and 0.96), $q$ decreases as $h$ increases, 
flattens out, bends over, and then drops discontinuously in a first--order 
transition to the P phase. 
At $\alpha_1=0.93$ (apparently corresponding to lower $T$ than used in 
Ref.\onlinecite{schotte98}), $q$ forms a reasonable plateau before rounding 
and falling to zero in a weakly first--order transition to the C phase; 
at slightly larger $h$, a second--order transition occurs to the P phase.  

\noindent 
Theoretically, the plateau in $q/q_0$ occurs at $\approx0.6$, almost 
independent of $\alpha_1$; the level of the plateau is reasonably robust 
(for $\gamma_2=0.3$, the plateau occurs at $q/q_0\approx0.56$). 
Figure 1 of the preceding article\cite{schotte98} compares these results with 
the available data. 
Experiment finds a plateau (as in theory), at about the theoretical level 
($q/q_0\approx0.6$), for both $T=10.34\,$K and $T=9.95\,$K; the latter data 
are slightly rounded at higher $H$, as in theory for $\alpha_1\alt0.98$. 
Data were not obtained at low fields where $q$ descends from the zero--field 
value to the plateau, preventing more detailed comparison with theory. 

Order parameters: 
Figure 3 shows the order parameter for the IC$_1$ phase, for Landau 
parameters $\alpha_1=0.96$ and $h=0.4$ (near the middle of the plateau 
in Figure 2). 
At maximum $m_x$, the configuration is almost colinear.  
Figure 4 shows the order parameter for the IC$_2$ phase, for 
$\alpha_1=1.005$ and $h=0.18$ (near the middle of the pocket in Figure 1). 

\noindent
The preceding article\cite{schotte98} finds qualitative agreement between 
theory and experiment for both IC phases, but it is not possible to determine 
the order parameters uniquely from experiment. 
The theoretical order parameter in the IC$_1$ phase is too distorted 
(likely because of the loop at the right--hand side in Figure 3); 
that in the IC$_2$ phase agrees reasonably well, although the third sublattice 
is not visible in the available data. 

\section{Summary} 

Landau theory with the biquadratic term (representing some effects of 
fluctuations) explains the appearance of the new IC phase 
found\cite{werner97,schotte98} near $T_{\rm N}$. 
The new phase is the Landau--theory counterpart of the 110 state studied 
in Ref.\onlinecite{jacobs93}, but stabilized by fluctuations. 
Only coarse adjustment of the only available parameter (the coefficient 
$\gamma_2$ of the biquadratic term) is needed to obtain qualitative 
(in some cases quantitative) agreement with experiment. 

In more detail: 
Landau theory finds a new IC phase to exist near $T_{\rm N}$. 
It explains\cite{schotte98} qualitatively the neutron--diffraction results 
in both IC phases. 
It does not explain the increase of $T_{\rm N}$ with field, but it explains 
other features of the phase diagram. 
It predicts moderately well the order of the transitions. 
It predicts that the wavenumber $q$ of the IC$_2$ phase is larger than in the 
IC$_1$ phase, as observed; the experimental $T$ dependence of $q$ is however 
weaker than predicted. 
It predicts that the plateau in the wavenumber of the IC$_1$ phase occurs 
at $q/q_0\approx0.6$, as observed. 
Theory and experiment agree qualitatively with respect to the order 
parameters. 

\acknowledgments 
We are grateful to U. Schotte and M. E. Zhitomirsky for discussions and to 
U. Schotte for informing us of the experimental results prior to publication. 
This research was supported by the Natural Sciences and Engineering Research 
Council of Canada and by the Japan Society for the Promotion of Science. 

\appendix 
\section{Extended Landau theory}

The N\'eel temperature $T_{\rm N}$ increases initially with field for the 
TAFM because of thermal fluctuations\cite{lee84,kawamiya84,miya86JPSJ}. 
It increases also for CsCuCl$_3$ (Refs.\onlinecite{werner97,schotte98}), for 
(one believes) the same reason. 
Since Equation (2) cannot explain the increase, it is natural to ask whether 
an extended Landau theory can do so. 
To investigate this question, we add the following fourth--order 
invariants\cite{plumer94} to the square brackets in Equation (2): 
$${\textstyle{1\over4}}\,\gamma_3 \,
              {\bf m}_{jl}^2 \sum_{k=1}^3 {\bf m}_{kl}^2
+\gamma_4\,{\bf m}_{jl}^2 \sum_{k=1}^3 {\bf m}_{kl}\cdot{\bf m}_{k+1,l} $$
\begin{equation} 
+{\textstyle{1\over2}}\,\gamma_5\,{\bf m}_{jl}\cdot{\bf m}_{j+1,l}
                  \sum_{k=1}^3 {\bf m}_{kl}\cdot{\bf m}_{k+1,l}
+         \gamma_6\,{\bf m}_{jl}^2 
            \left({\bf m}_{j-1,l}\cdot{\bf m}_{j+1,l}\right) \ ;
\end{equation}
our coefficients are related to those of Ref.\onlinecite{plumer94} by 
$$ B_1=\gamma_1-{\textstyle{1\over2}}\gamma_2+3\gamma_3
-6\gamma_4+{\textstyle{3\over2}}\gamma_5 -2\gamma_6\ ,$$
$$B_2=\gamma_1-2\gamma_2+4\gamma_6\ ,$$
$$B_3=\gamma_1-2\gamma_2+3\gamma_3+12\gamma_4+6\gamma_5-4\gamma_6\ ,$$
$$B_4=\gamma_1-{\textstyle{1\over2}}\gamma_2 -2\gamma_6\ ,$$ 
$$B_5=\gamma_1+\gamma_2+3\gamma_3+3\gamma_4-3\gamma_5+\gamma_6\ ,$$
\begin{equation} 
B_6=\gamma_1+\gamma_2+\gamma_6\ .
\end{equation}

For the stacked TAFM, at $T$ sufficiently below $T_{\rm N}$, the phase 
sequence with decreasing $H$ must be: 
P phase $\to$ C phase with ${\bf m}_1={\bf m}_2$ $\to$ colinear C phase 
$\to$ low--field C phase; 
a little analysis gives the requirements $\gamma_2>0$, $B_3>0$ and $B_5>0$. 
If the P$\to$C transition is second--order (as at $h=0$, $\alpha_1=\alpha_2$), 
we find that the phase boundary is given by 
\begin{equation} 
h^2=\left(\frac{\alpha_2-\alpha_1}{B_5}\right)
\left[(\alpha_1+2\alpha_2)+\frac{B_3}{B_5}(\alpha_2-\alpha_1) \right]^2\ .
\end{equation}
Since $B_5>0$, the extended Landau theory cannot explain this effect of 
fluctuations (the increase of $T_{\rm N}$ with $H$) for the stacked TAFM; 
therefore we believe that it cannot explain either the increase of 
$T_{\rm N}$ with $H$ for CsCuCl$_3$. 

We also used the extended theory to determine several phase diagrams like 
Figure 1, for several different sets of parameters. 
The C, P and IC$_1$ phases always appear, as does the IC$_2$ phase 
(unless of course $\gamma_2$ is sufficiently negative). 
Generally, the more complicated free energies (those with more of the 
$\gamma_i$ parameters $\neq0$) give more complicated phase diagram (with for 
example re--entrant phases), but no new phases are found.

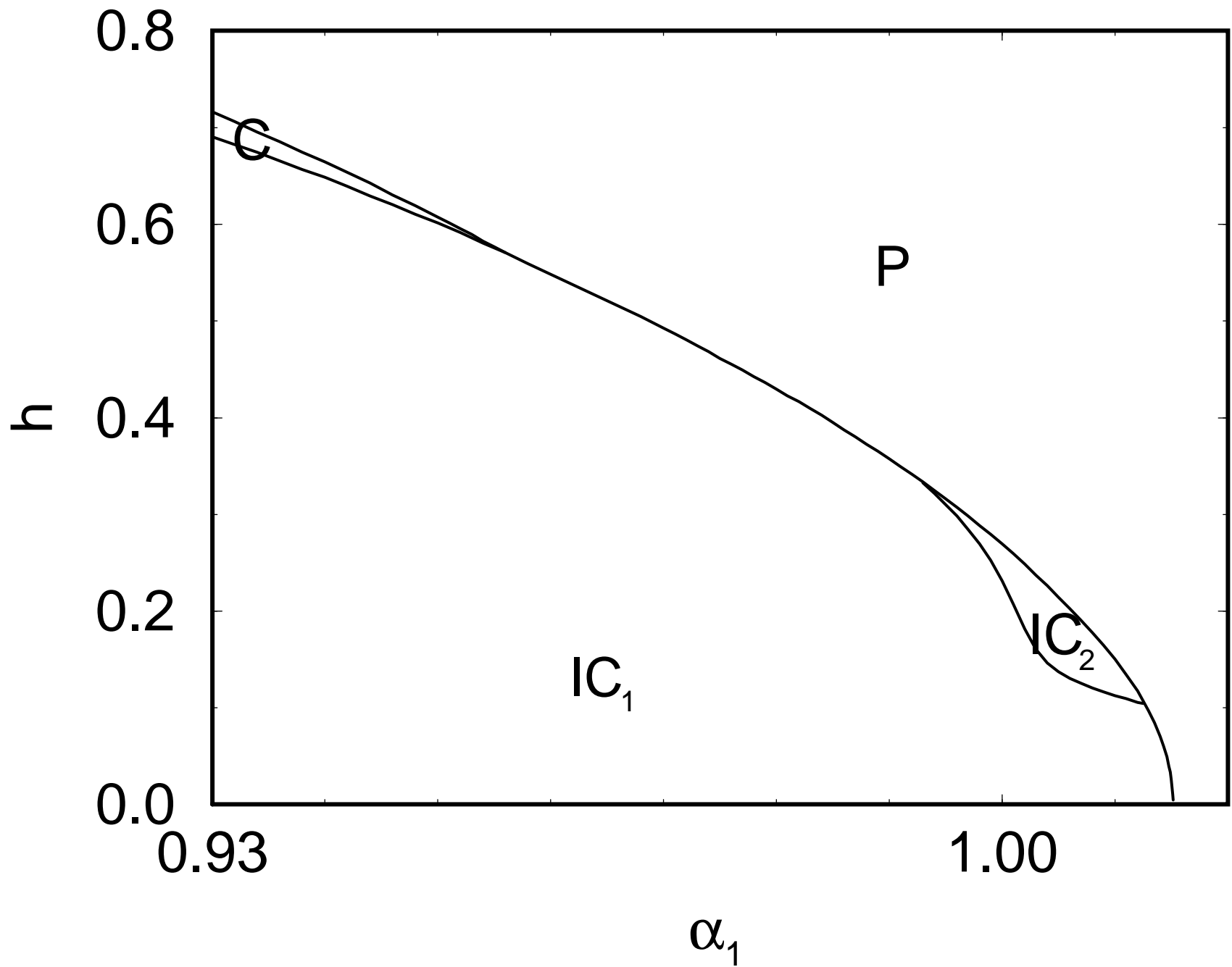
\begin{figure} 
\caption{
Phase diagram in the $\alpha_1$--$h$ plane for $\gamma_2=0.2$. 
The Landau parameter $\alpha_1$ is linear in the temperature $T$, and 
$h\approx0.88H/H_{\rm S}(0)$. 
The paramagnetic phase (P), the commensurate phase (C), and the two 
incommensurate phases (IC$_1$ and IC$_2$) are optimal in the regions 
indicated. 
}
\end{figure}

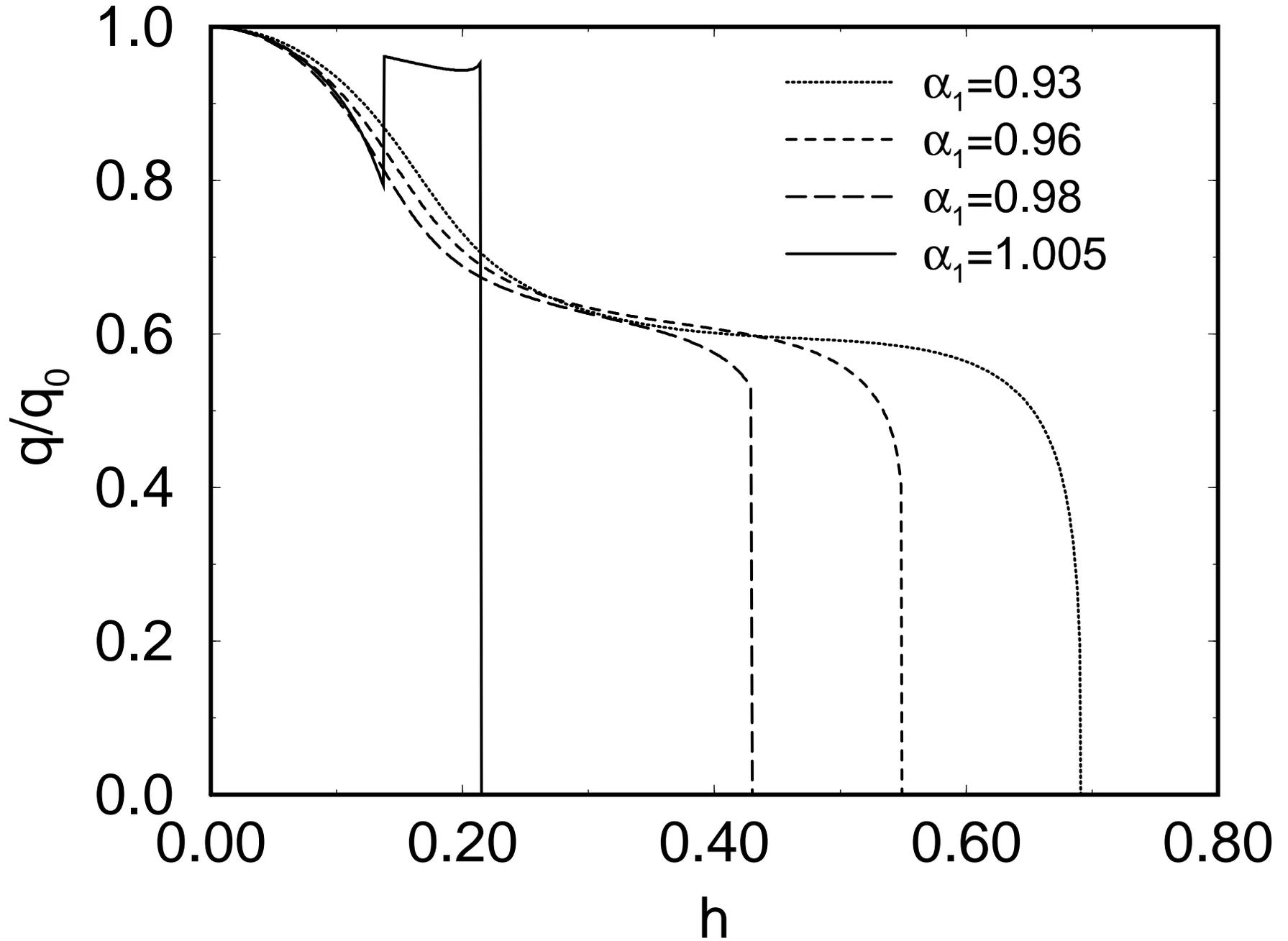
\begin{figure} 
\caption{
Reduced wavenumber $q/q_0$ as a function of the Landau field $h$ for 
four values of the Landau parameter $\alpha_1$. 
}
\end{figure}

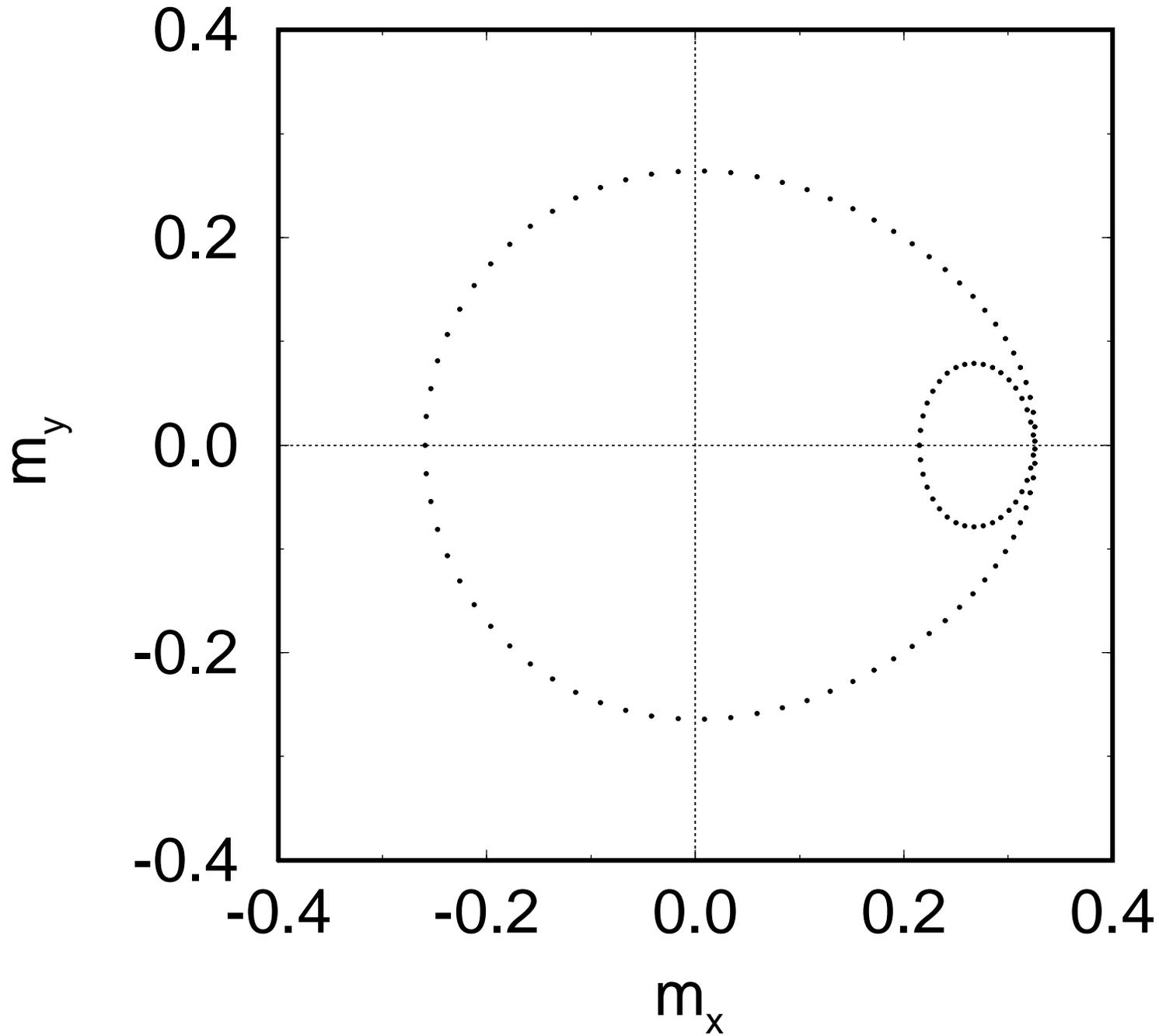
\begin{figure} 
\caption{
IC$_1$ phase: orbit in the $m_x$--$m_y$ plane for the order parameter on one 
of the three equivalent sublattices. 
The Landau parameters are $\alpha_1=0.96$ and $h=0.4$ (near the 
middle of the plateau in Figure 2); the period $L$ is 116. 
The order parameters on the other two sublattices are displaced by 
$l=\pm L/3$ from the first. 
} 
\end{figure}

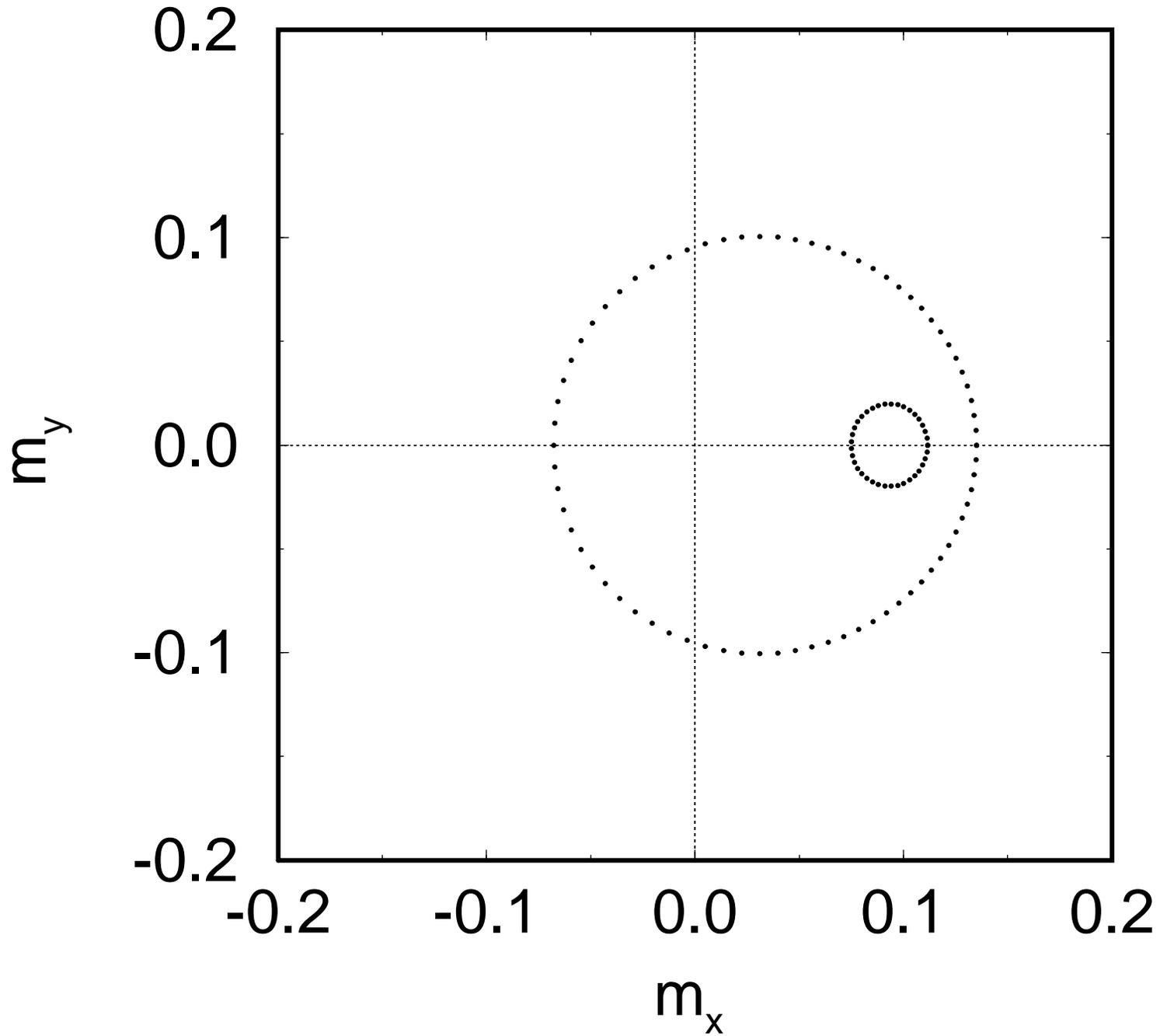
\begin{figure} 
\caption{
IC$_2$ phase: orbits in the $m_x$--$m_y$ plane for the order parameters on 
two of the three sublattices. 
The Landau parameters are $\alpha_1=1.005$ and $h=0.18$ (near the 
middle of the pocket in Figure 1); the period $L$ is 74. 
The outer loop is the orbit for one of the two equivalent sublattices; 
the orbit for the second is displaced by $l=L/2$. 
The order parameter on the third sublattice (inner loop) wobbles about the 
field with period $L/2$. 
} 
\end{figure}


\begin{references}

\bibitem[*]{Allan} Electronic address: jacobs@physics.utoronto.ca

\bibitem[\dag]{Tetsuro} Address after March 1998: 
Department of Physics, Tokyo Institute of Technology, Oh--okayama, 
Meguro--ku, Tokyo, Japan 152; nick@stat.phys.titech.ac.jp 

\bibitem{collins97} Collins M F and Petrenko O A 
1997 {\it Can. J. Phys.} {\bf75} 605 

\bibitem{achiwa69} Achiwa N 
1969 {\it J. Phys. Soc. Jpn.} {\bf27} 561 

\bibitem{moto78} Motokawa M 1978 unpublished paper presented at the Annual 
Meeting of the Physical Society of Japan 

\bibitem{DM} Dzyaloshinskii I 1958 {\it J. Phys. Chem. Solids} {\bf4} 241; 
Moriya T 1960 {\it Phys. Rev.} {\bf120} 91 

\bibitem{adachi80} Adachi K Achiwa N and Mekata M 
1980 {\it J. Phys. Soc. Jpn.} {\bf49} 545 

\bibitem{lee84} Lee D H Joannopoulos J D Negele J W and Landau D P 
1984 {\it Phys. Rev. Lett.} {\bf52} 433; 
1986 {\it Phys. Rev. B} {\bf33} 450 

\bibitem{kawa84} Kawamura H 
1984 {\it J. Phys. Soc. Jpn.} {\bf53} $2\,452$; 
1991 {\it J. Phys. Soc. Jpn.} {\bf61} $1\,299$ 

\bibitem{kawamiya84} Kawamura H and Miyashita S 
1984 {\it J. Phys. Soc. Jpn.} {\bf53} $4\,138$; 
1985 {\it J. Phys. Soc. Jpn.} {\bf54} $4\,530$. 
Miyashita S and Kawamura H 
1985 {\it J. Phys. Soc. Jpn.} {\bf54} $3\,385$ 

\bibitem{korsh86} Korshunov S E 
1986 {\it J. Phys. C: Solid State Phys.} {\bf19} $5\,927$ 

\bibitem{nishi86} Nishimori H and Miyashita S 
1986 {\it J. Phys. Soc. Jpn.} {\bf55} $4\,448$ 

\bibitem{chub91} Golosov D I and Chubukov A V 
1989 {\it Pis'ma Zh. Eksp. Teor. Fiz.} {\bf50} 416 
(1990 {\it JETP Lett.} {\bf50} 451); 
Chubukov A V and Golosov D I 
1991 {\it J. Phys.: Condens. Matter} {\bf3} 69 

\bibitem{fed85} Fedoseeva N V Gekht R S Velikanova T A and Balaev A D 
1985 {\it JETP Lett.} {\bf41} 406 

\bibitem{nojiri88} Nojiri H Tokunaga Y and Motokawa M 
1988 {\it J. Phys. (Paris) Colloq.} {\bf49} C8--$1\,459$ 

\bibitem{shiba93} Shiba H and Nikuni T 
1993 {\it Recent Advances in Magnetism of Transition Metal Compounds}, 
edited by Kotani A and Suzuki N (World Scientific, Singapore) p. 372 

\bibitem{nikuni93} Nikuni T and Shiba H 
1993 {\it J. Phys. Soc. Jpn.} {\bf62} $3\,268$ 

\bibitem{tanaka92} Tanaka H Schotte U and Schotte K--D 
1992 {\it J. Phys. Soc. Jpn.} {\bf61} $1\,344$ 

\bibitem{ohta93} Ohta H Imagawa S Motokawa M and Tanaka H 
1993 {\it J. Phys. Soc. Jpn.} {\bf62} $3\,011$ 

\bibitem{chiba94} Chiba M Ajiro Y and Morimoto T 
1994 {\it Physica B} {\bf201} 200 

\bibitem{mino94} Mino M Ubukata K Bokui T Arai M Tanaka H and Motokawa M 
1994 {\it Physica B} {\bf201} 213 

\bibitem{ohta94} Ohta H Imagawa S Motokawa M and Tanaka H 
1994 {\it Physica B} {\bf201} 208 

\bibitem{schotte94} Schotte U St\"u{\ss}er N Schotte K--D 
Weinfurter N Mayer H M and Winkelmann M 
1994 {\it J. Phys.: Condens. Matter} {\bf6} $10\,105$ 

\bibitem{chiba95} Chiba M Ajiro Y and Morimoto T 
1995 {\it J. Magn. Magn. Mater.} {\bf140}--{\bf144} $1\,673$ 

\bibitem{moto95} Motokawa M and Arai M 
1995 {\it Physica B} {\bf213}--{\bf214} $1\,017$ 

\bibitem{weber96} Weber H B Werner T Wosnitza J v L\"ohneysen H and Schotte U 
1996 {\it Phys. Rev. B} {\bf54} $15\,924$ 

\bibitem{jacobs93} Jacobs A E Nikuni T and Shiba H 
1993 {\it J. Phys. Soc. Jpn.} {\bf62} $4\,066$ 

\bibitem{ohyama95} Ohyama T and Jacobs A E 
1995 {\it Phys. Rev. B} {\bf52} $4\,389$ 

\bibitem{nojiri97} Nojiri H Takahashi K Fukuda T Fujita M Arai M 
and Motokawa M 
1997 paper presented at the International Conference on 
Neutron Scattering, Toronto, August 

\bibitem{werner97} Werner T Weber H B Wosnitza J Kelnberger A Meschke M 
Schotte U St\"u{\ss}er N Ding Y and Winkelmann M 
1997 {\it Solid State Commun.} {\bf102} 609 

\bibitem{nikuni98} Nikuni T and Jacobs A E 
1998 {\it Phys. Rev. B} {\bf57} ? (LANL preprint number 9702201) 

\bibitem{miya86JPSJ} Miyashita S 
1986 {\it J. Phys. Soc. Jpn.} {\bf55} $3\,605$ 

\bibitem{schotte98} Schotte U Kelnberger A and St\"u{\ss}er N 
1998 preceding article 

\bibitem{hyodo81} Hyodo H Iio K and Nagata K 
1981 {\it J. Phys. Soc. Jpn.} {\bf50} $1\,545$ 

\bibitem{tazuke81} Tazuke Y Tanaka H Iio K and Nagata K 
1981 {\it J. Phys. Soc. Jpn.} {\bf50} $3\,919$ 

\bibitem{mekata95} Mekata M Ajiro Y Sugino T Oohara A Ohara K 
Yasuda S Oohara Y and Yoshizawa H 
1995 {\it J. Magn. Magn. Mater.} {\bf140}--{\bf144} $1\,987$ 

\bibitem{jensen} Jensen J, unpublished 

\bibitem{plumer94} Plumer M L Mailhot A and Caill\'{e} A 
1994 {\it Phys. Rev. B} {\bf49} $15\,133$ 

\bibitem{plumer97} Plumer M L and Mailhot A 
1997 {\it J. Phys.: Condens. Matter} {\bf9} L165 

\bibitem{schotte98b} Schotte U private communication
\end{references}
\end{document}